\documentclass[vci-paper]{vciart}
\usepackage{amssymb}
\usepackage{graphics,graphicx,epsfig}
\usepackage{epstopdf}
\begin{document}

\begin{frontmatter}

\title{Detecting Double Beta Decays Using Nuclear Emulsions}

% if there is only one institution, use this form:
%\author{John Author, Giovanna Autore}
%\address{University of Wisdom, Physics City, Scienceland}

% else, use optional labels to link authors explicitly to addresses,
% as shown below:
\author[ires]{M.~Dracos\thanksref{contact}},

\address[ires]{IPHC, Universit\'{e} Louis Pasteur, CNRS/IN2P3, Strasbourg, France}

\thanks[contact]{Corresponding author, marcos.dracos@ires.in2p3.fr}

\begin{abstract}

Neutrino nature and absolute mass scale are major questions in particle physics which cannot be addressed by the present neutrino oscillation program.
To answer these two questions, several neutrinoless double beta decay experiments are underway or planed for the near future.
These experiments, mainly use bolometric techniques or gaseous counters coupled with scintillator detectors.
The energy resolution is better in bolometric experiments but experiments coupling tracking with calorimetry have the advantage of observing the two electron tracks and remove many background sources.
Here, we present a proposal of using nuclear emulsions to observe double beta decays.
This technique has the advantage of precise tracking and vertexing even for low energy electrons.

\end{abstract}

\end{frontmatter}

\section{Introduction}

In order to prove that neutrinos are Majorana particles (neutrinos are identical with anti--neutrinos) and measure the effective neutrino mass, several experiments try to observe neutrinoless double beta decays (see for example~\cite{cuoricino,nemo3,gerda,snemo,cuore,majorana}).
For this proposal of using nuclear emulsions to detect neutrinoless double beta decays, the running NEMO3 experiment~\cite{nemo3} and the proposed one Super--NEMO~\cite{snemo}, will be used as reference.

\begin{figure}[hbt]
\begin{center}
\epsfig{file=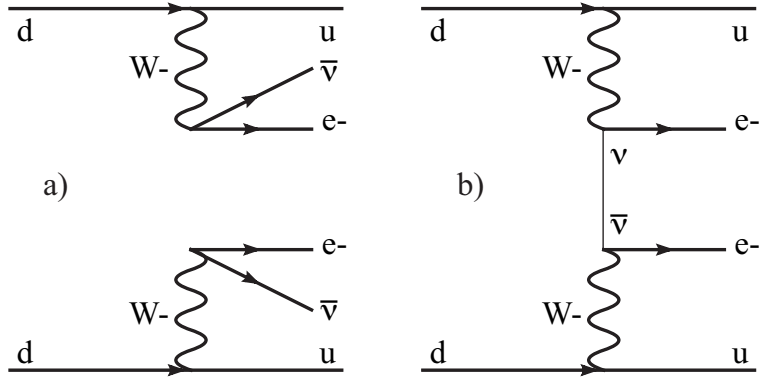, width=10cm}
\caption{\small Allowed (a) and forbidden (b) double beta decay.} \label{double_beta}
\end{center}
\end{figure}

Double beta decay is of two types, the allowed one already observed, and the forbidden one not yet observed violating the leptonic number by two units.
Fig.~\ref{double_beta} depicts these two processes.
In the first one (allowed) there is emission of two electrons and two neutrinos while for the second one (forbidden) only two electrons are emitted carrying all the process energy. These energy (2.8--4.3~MeV) depends on the decaying nucleus.

\begin{figure}[hbt]
\begin{center}
\epsfig{file=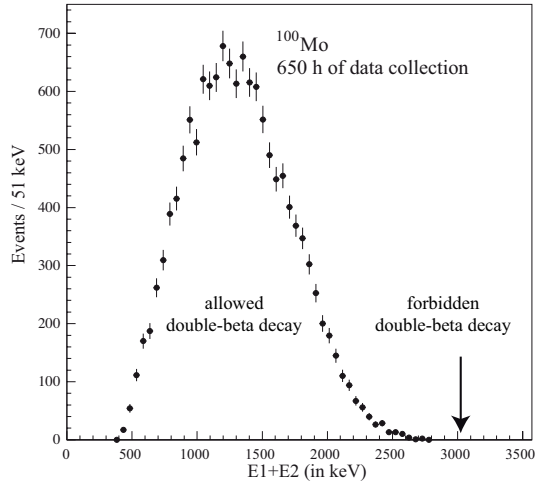, width=7cm}
\caption{\small Observed double--beta decay for Molybdenum by NEMO3 experiment.} \label{nemo1}
\end{center}
\end{figure}

In double--beta experiments the two electrons of both processes are detected.
For the first case, the electron energy distribution is quite broad due to the missing neutrino energy while for the forbidden process the energy is well defined mainly dispersed by the detector energy resolution.
Fig.~\ref{nemo1} presents the observed energy distribution of the two electrons by NEMO3 experiment for $^{100}$Mo isotope. In this figure is also indicated the expected position of the energy peak corresponding to an eventual forbidden double--beta decay.

Two approaches have been adopted by these experiments, one consisting of just measuring with high precision the energy and a second one measuring the energy and also detecting the two electrons. The first one is affected by many other processes deposing energy in the detector ($\gamma$'s...) while the second one has less energy resolution but better background rejection. NEMO3 and the proposed Super--NEMO experiments are in this second category. NEMO3 utilise double--beta decaying isotopes in a foil form surrounded by gaseous detectors to detect the two electron tracks and scintillator detectors to measure their energy.

To achieve a neutrino effective mass limit of the order of 0.3~eV, NEMO3 experiments uses 7~Kg isotopes with a target foil surface of 20~m$^2$ and thickness of about
60~$\mu$m. The proposed Super--NEMO project aiming to reach an effective neutrino mass of 50~meV will use 10 times more isotope mass and also increase the energy resolution from 15\%  to less than 7\% for electrons at 1~MeV. To achieve these goals, Super--NEMO would use a modular detector for which each module will have a size of about $1\times 5\times 4$~m$^3$ for a total of 1500~m$^3$.

To obtain similar performances, nuclear emulsions could be used. OPERA experiment~\cite{opera} uses extensively this technique and considerable efforts have been done to automatise and accelerate the emulsion scanning.

\section{Description}

\begin{figure}[hbt]
\begin{center}
\epsfig{file=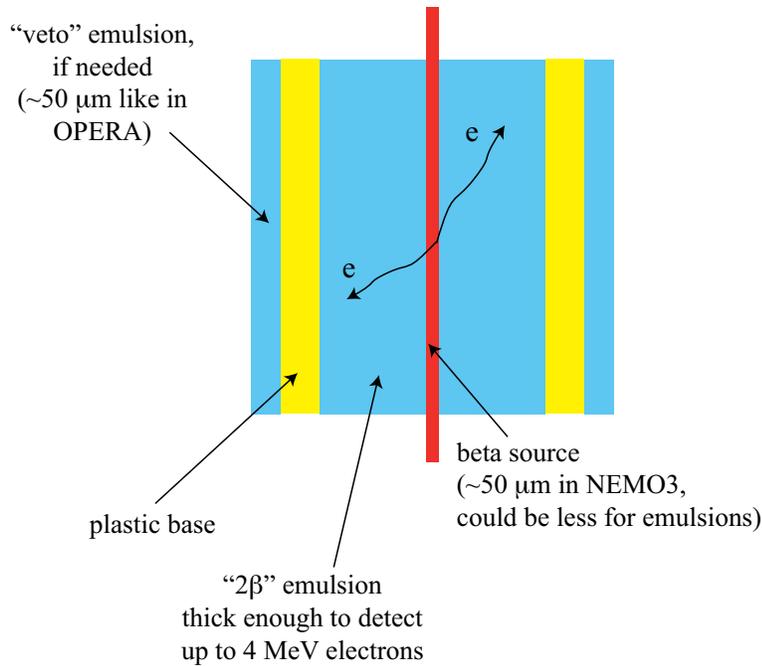, width=10cm}
\caption{\small Detection principle of of double--beta decays using emulsion sheets.} \label{principe}
\end{center}
\end{figure}

A basic unit of this double--beta emulsion detector consists of an isotope foil having at each side an emulsion sheet thick enough to stop up to about 3~MeV electrons (Fig.~\ref{principe} ) acting as tracker and calorimeter. At the outside part of the plastic base of the emulsions, a thin emulsion layer ($\sim 50\  \mu$m like OPERA emulsions) could be used to tag tracks coming from outside. The last $\sim 50\  \mu$m of the inner emulsion (detecting electrons from double--beta decays) could also be used to reject tracks coming from outside. Envelops of isotope/emulsions could be produced and put in boxes in a temperature controlled and low radioactivity environment. After a certain exposure time, the emulsions could be removed, developed and scanned in order to find electrons emerging from the same point of the isotope foils. The number of grains in the emulsion could give the electron energy. The removed isotope sheets could be used again with new emulsion sheets.
 
\begin{figure}[hbt]
\begin{center}
\epsfig{file=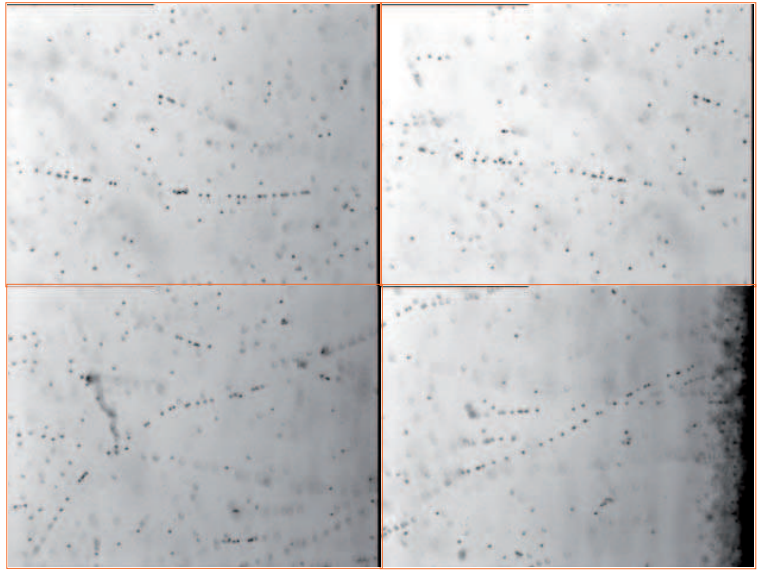, width=10cm}
\caption{\small 1~MeV electron tracks observed in OPERA nuclear emulsions.} \label{ariga_electrons}
\end{center}
\end{figure}

This experimental technique must be able to detect electrons with a kinetic energy as low as 1~MeV and even lower. Fig.~\ref{ariga_electrons} presents 1~MeV electron tracks observed by Nagoya University using a $^{90}$Sr source and OPERA emulsions~\cite{ariga}. These tracks are well visible and their number of grains can be counted.

\section{Experiment}

NEMO3 isotope surface is 20~ m$^2$ while Super-NEMO will use a surface of $10\times 20$~m$^2$. 
To cover the same isotope source surface with emulsions (both sides to detect the 2 electrons) an emulsion surface of $2\times 200=400$~m$^2$ is needed.
Just for comparison, one OPERA emulsion has about $0.012$~m$^2$ and one brick 0.680~m$^2$.
Thus, 400~m$^2$ is about the equivalent of 600 OPERA bricks but of course not with the same thickness (OPERA has more than 150000 bricks).
These emulsions could be stored in a few cubic meters underground room.

The same envelops like the OPERA changeable sheets could be used by introducing at the middle of the two emulsions a double beta source sheet, or use longer emulsion sheets easier to handle by microscopes.
The requirement of having thick emulsions (more than 2~mm) could probably limit the emulsion sheet surface. 
These envelops could be kept for some time (e.g. 6--12 months depending on fading) in the experiment and after this period they could be removed, developed and scanned periodically one after the other.
They could be replaced by new envelops (recycling the isotopes) during 5 years in order to accumulate something equivalent to what Super-NEMO could do: $\sim 400\times 5$ year$\times$m$^2$.

An issue could be the time needed to make a full scan of 2000~m$^2$.
It is possible that a full scan of the whole emulsion volume will not be needed.
If for example in the first $50\  \mu$m from the isotope surface nothing has been found, it would not be necessary to go deeper in the emulsion. 
If the Japanese S-UTS scanning system developed by Nagoya University is used with a speed of 50~cm$^2$/hour, for one scanning table, 25~m$^2$/year (200 working days/year) have to be scanned.
By using 16 tables and extracting 100~m$^2$ every 3 months (1 year exposure at the beginning and putting back new emulsions with the same isotopes), this finally will take less than 5  years (usual duration of double beta experiments).
Probably, the emulsion thickness needed to detect these electrons will need more scanning time than OPERA ones and the speed would be significantly less than 50~cm$^2$/h.
On the other hand, scanning speed increases very fast with time (see T. Nakano's presentation during this same workshop).

\section{Pending questions}

A very important parameter to know is the energy resolution which could be achieved by this technique.
The energy resolution for 1~MeV electrons could be obtained using thick emulsions (more than 1~mm) by irradiation using $^{207}$Bi source providing 976~MeV electrons. Algorithms have to be developed to recognise these low energy tracks and attribute to them the right grains. The aim is to obtained an energy resolution better than 7\%. A magnetic field could probably help to increase the energy resolution but this could do the experiment heavier to realise.
The possibility to take thinner isotope sheets than NEMO3/Super--NEMO could also help to have better energy resolution.

The reconstruction efficiency for NEMO3 is of the order of 15\% while the required one for Super--NEMO is higher than 40\%.
These experiments have an electron energy cut at 0.2~MeV.
The emulsion experiment would probably not be able to detect electrons with an energy lower than 1~MeV.
In order to have similar efficiency than Super--NEMO, this experiment must have better than 40\% reconstruction efficiency or increase the isotope quantity.

Another important parameter will be the affordable background density.
Again, for this R\&D, a  $^{207}$Bi source of well known activity could be used.
Emulsions could be irradiated during a specific time on a certain surface.
After that, one could try to reconstruct and count the associated electron tracks in order to estimate the reconstruction efficiency (the energy distribution could be compared with the expected one).
This parameter could also be estimated by simulation.

\begin{figure}[hbt]
\begin{center}
\epsfig{file=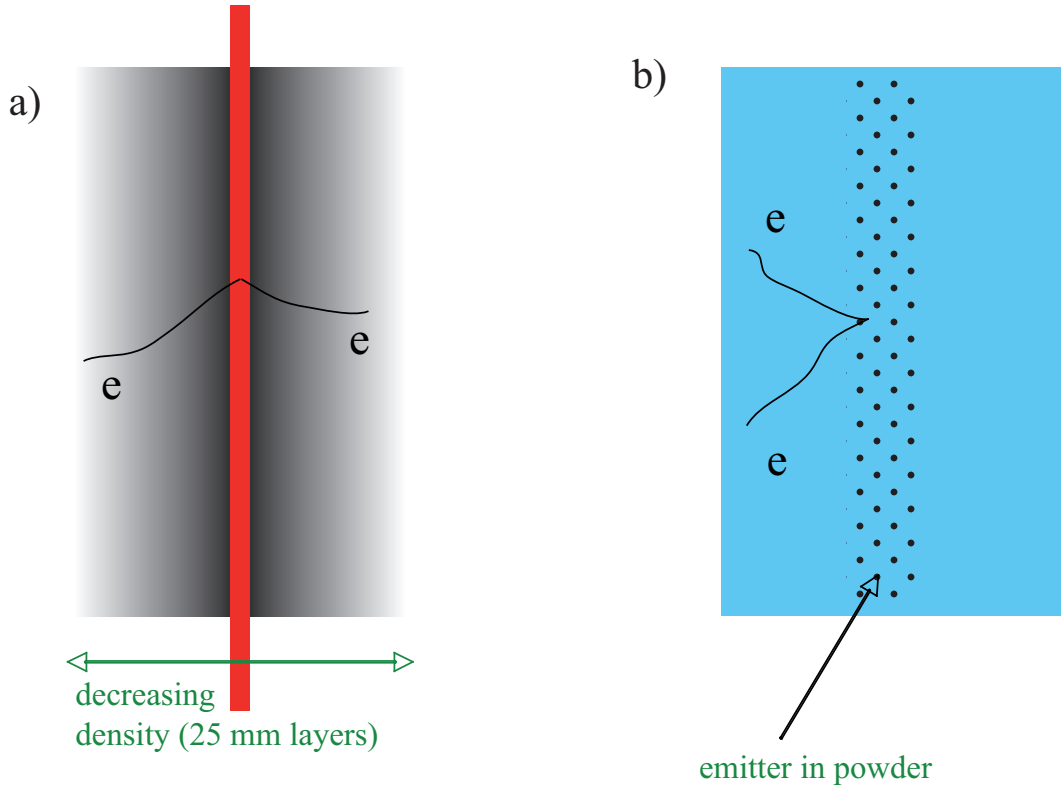, width=14cm}
\caption{\small Emulsion sheets with a) variable grain density to optimise the emulsion thickness and b) isotopes in powder form deluded at the centre of the emulsion layer for better vertex and energy reconstruction.} \label{ideas}
\end{center}
\end{figure}

\section{Conclusion}

An emulsion based double beta experiment has been investigated.
This experiment could be very compact, benefit from OPERA scanning developments and give similar results than other proposed experiments.
R\&D is needed to determine the energy resolution, the reconstruction efficiency and the background which could be afforded by this technique.

\section{Acknowledgement}

I'm very grateful to my Nagoya University OPERA colleagues, especially to M. Nakamura and A. Ariga for all fruitful discussions about emulsion and scanning techniques. I would also like to thank R. Arnold for the valuable discussions we had concerning double beta decays and possible backgrounds.

%\clearpage

\end{document}